\documentclass{optica-article}

\journal{opticajournal} 

\articletype{Research Article}

\usepackage{amsmath}
\usepackage{multirow}

\newcommand\mat\mathbf
\newcommand\dd{\, \mathrm{d}}
\newcommand\ii{\mathrm{i}}

\newcommand\um{~{\mu}\textrm{m}}
\newcommand\nm{\textrm{~nm}}
\newcommand\mm{\textrm{~mm}}
\renewcommand\Im{\operatorname{Im}}
\renewcommand\Re{\operatorname{Re}}

\begin{document}

\title{Design of diffractive neural networks solving different classification problems at different wavelengths}

\author{Georgy~A.~Motz,\authormark{1,2} 
Leonid~L.~Doskolovich,\authormark{1,2,*} 
Daniil~V.~Soshnikov,\authormark{1,2}
Egor~V.~Byzov,\authormark{1,2}
Evgeni~A.~Bezus,\authormark{1,2}
Nikita~V.~Golovastikov,\authormark{1,2}
and Dmitry~A.~Bykov\authormark{1,2}
}

\address{\authormark{1}Samara National Research University, 34 Moskovskoye shosse, Samara 443086, Russia\\
\authormark{2}Image Processing Systems Institute, National Research Centre ``Kurchatov Institute'', 151 Molodogvardeyskaya st., Samara 443001, Russia}
						
\email{\authormark{*}leonid@ipsiras.ru}

\begin{abstract*}
We consider the problem of designing a diffractive neural network (DNN) consisting of a set of sequentially placed phase diffractive optical elements (DOEs) and intended for the optical solution of several given classification problems at different operating wavelengths, so that each classification problem is solved at the corresponding wavelength.
The problem of calculating the DNN is formulated as the problem of minimizing a functional that depends on the functions of the diffractive microrelief height of the DOEs constituting the DNN and represents the error in solving the given classification problems at the operating wavelengths. 
We obtain explicit and compact expressions for the derivatives of this functional and, using them, formulate a gradient method for the DNN calculation.
Using this method, we design DNNs for solving the following three classification problems at three different wavelengths: the problem of classifying handwritten digits from the MNIST database, the problem of classifying fashion products from the Fashion MNIST database, and the problem of classifying ten handwritten letters from the EMNIST database. 
The presented simulation results of the designed DNNs demonstrate high performance of the proposed method.
\end{abstract*}

\section{Introduction}
In recent years, the design of photonic structures for optical computing and optical information processing has attracted great interest.
These structures are considered as a promising platform for the further development of computing systems and are intended for creating an alternative to electronic components or supplementing them~\cite{1,2,3,4}.
Optical neural networks~\cite{5,6,7,8,9}, and, in particular, diffractive neural networks (DNNs) comprising a cascade of sequentially placed phase diffractive optical elements (DOEs)~\cite{10,11,12,13,14,15,16,17,18,19,20,21,22,23,24,25,26} are considered as one of the most promising and rapidly developing areas in the field of optical information processing.
It should be noted that DOEs (both single and cascaded) have a long history and are widely used for solving a large class of problems of steering laser radiation~\cite{27,28,29,30,31,32,33}.
At the same time, the use of cascaded DOEs for optical solution of machine learning problems was first demonstrated only in 2018 in the seminal paper~\cite{10}.
In this work, the authors pointed out several analogies between a cascade of DOEs and ``conventional'' artificial neural networks and introduced the term ``diffractive deep neural network''.
The possibility of the optical solution of classification problems using cascaded DOEs was theoretically and experimentally demonstrated in~\cite{10}. 
Subsequent works considered the use of DNNs (cascaded DOEs) for solving various classification problems~\cite{11,12,13,14,15,17,25,26}, 
object and video recognition~\cite{13, 15}, 
salient object detection~\cite{11}, 
implementing multispectral imaging~\cite{22}, 
and performing matrix multiplication, 
as well as implementing other linear operators~\cite{12, 18, 20, 21}.
The main method for designing DNNs is the stochastic gradient descent method, as well as ``improved'' first-order methods based on it~\cite{34}. 
These methods have become widely used and have shown their high efficiency in beam shaping problems traditionally solved using DOEs~\cite{35,36}.

In most works, DNNs are calculated to work with radiation of a single operating wavelength.
At the same time, the problem of calculating DNNs designed to work with radiation of various wavelengths is of great scientific and practical interest.
In what follows, we will refer to such DNNs as spectral DNNs (or cascaded spectral DOEs). 
Spectral DNNs can be used to process spectral data, carry out parallel computations by simultaneously solving several machine learning problems at different wavelengths, change their functionality (i.\,e., the problem being solved) depending on the wavelength of the incident radiation, etc.
In particular, the works~\cite{23, 24} considered the calculation of DNNs for spectral filtering and spectral analysis of the incident radiation.
In~\cite{21, 22}, spectral DNNs were considered for the optical implementation of various linear transformations at different wavelengths (each transformation being carried out at its ``own'' wavelength), as well as for multispectral imaging.
One of the main problems, in which DNNs have demonstrated high efficiency, is the problem of optical image classification.
In the opinion of the present authors, the problem of calculating spectral DNNs for solving several different classification problems at different wavelengths has not yet been sufficiently studied.
In particular, although the solution of classification problems using the radiation of several different wavelengths was considered in recent works~\cite{25, 26}, several wavelengths were used only to improve the quality of the solution of a single fixed classification problem.
Thus, the solution of several different classification problems at different wavelengths has not been considered in~\cite{25, 26} and, to the best of our knowledge, in the other existing works.

In this work, we consider the design of spectral DNNs (cascaded spectral DOEs) for solving several different classification problems at several different wavelengths. 
We formulate the problem of calculating a spectral DNN as the problem of minimizing a functional representing the error of solving the given classification problems at the operating wavelengths. 
This functional depends on the functions defining the diffractive microrelief height of the DOEs constituting the DNN.
Explicit and compact expressions are obtained for the derivatives of the error functional and on this basis, a gradient method for the DNN design is presented.
Using the proposed gradient method, we calculate several examples of spectral DNNs for solving the following three problems: 
classification of handwritten digits from the MNIST database at the wavelength of 457~nm, 
classification of fashion products from the Fashion MNIST database at 532~nm, 
and classification of ten handwritten letters from A to J (lowercase and uppercase) from the EMNIST database at 633~nm. 
The presented numerical simulation results demonstrate good classification accuracy provided by the designed spectral DNNs.

\section{Design of spectral DNNs for solving several classification problems}\label{sec:1}

\begin{figure}[hbt]
	\centering
		\includegraphics{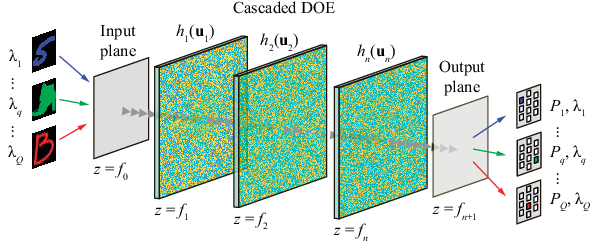}
	\caption{\label{fig:1} Geometry of the problem of calculating a DNN for solving different classification problems at different wavelengths.}
\end{figure}

Let us consider the problem of calculating a spectral DNN (a cascaded DOE) intended for solving several classification problems $P_q, q = 1,\ldots ,Q$ at different wavelengths $\lambda_q,\,\,q = 1,\ldots ,Q$, so that each classification problem $P_q$ is solved at the corresponding wavelength $\lambda_q$.
We assume that the cascaded DOE consists of $n$ phase DOEs located in the planes $z = f_1, \ldots, z = f_n$ $(0 < f_1 < \cdots < f_n)$ and defined by the functions of diffractive microrelief height $h_1( \mat{u}_1 ),\ldots, h_n( \mat{u}_n )$, where $\mat{u}_j = ( u_j, v_j )$ are Cartesian coordinates in the planes $z = f_j$ (Fig.~\ref{fig:1}).

Let us first describe the required operation of the DNN at a certain single wavelength $\lambda_q$.
We assume that in the input plane $z=0$, amplitude images of objects from ${N_q}$ different classes corresponding to the classification problem $P_q$ are sequentially generated.
Each generated image is illuminated by a plane wave with wavelength $\lambda_q$.
Let us denote by $w_{0,q,j}(\mat{u}_0 )$ the complex amplitude of the light field generated in this way in the input plane. 
Here and in what follows, the subscript of a certain complex amplitude of the field $w_{m,q,j}(\mat{u}_m )$ contains 
the index $m$ of the plane, in which this amplitude is defined, 
the wavelength index $q$ (which is also the index of the corresponding classification problem), and 
the class number $j$ of the input image.

The light field $w_{0,q,j}(\mat{u}_0 )$ generated at $z=0$ then propagates through the cascaded DOE to the output plane $z = f_{n+1}$.
We assume that the light propagation in the free space (between the planes, in which the DOEs are located) is described by the Fresnel--Kirchhoff diffraction integral, and that the transmission of the light field through a DOE can be described in the thin optical element approximation as the multiplication of the beam complex amplitude by the complex transmission function (CTF) of this DOE.
The CTF of the $m$-th DOE is wavelength-dependent and for the wavelength $\lambda_q$ has the form
\begin{equation}
\label{eq:1}
{T_{m,q}}(\mat{u}_m ) = \exp \left\{ \ii \varphi_{m,q}(\mat{u}_m ) \right\} = \exp \left\{ \ii\frac{2\pi }{\lambda_q} [ n(\lambda_q) - 1] h_m ( \mat{u}_m ) \right\},
\end{equation}
where ${\varphi_{m,q}}(\mat{u}_m )$ is the phase function of the DOE (the phase shift introduced by the DOE) at the wavelength $\lambda_q$ and $n(\lambda_q)$ is the refractive index of the DOE material.
Under the made assumptions, the propagation of the input beam $w_{0,q,j}( \mat{u}_0 )$ from the input plane $z = 0$ through the cascaded DOE to the output plane $z = f_{n+1}$ is described by the following recurrent formula:
\begin{equation}
\label{eq:2}
\begin{aligned}
   w_{1,q,j} ( \mat{u}_1 ) 
	= 
	C_{q,1} \iint &w_{0,q,j}( \mat{u}_0 )\exp \left\{ \ii\frac{\pi}{\lambda_q d_1} ( \mat{u}_1  - \mat{u}_0 )^2 \right\}\dd^2\mat{u}_0 , 
  \\ w_{m,q,j}( \mat{u}_m ) 
	= 
	C_{q,m}\iint &w_{m - 1,q,j}( \mat{u}_{m-1} )T_{m - 1,q}( \mat{u}_{m-1} )
	\cdot\exp \left\{ \ii\frac{\pi }{\lambda_q d_m} ( \mat{u}_m - \mat{u}_{m-1} )^2 \right\}\dd^2\mat{u}_{m-1} ,
	\\&\hspace{10em}
	m = 2,\ldots ,n + 1, 
\end{aligned}
\end{equation}
where $w_{m,q,j}( \mat{u}_m ),\,m = 1, \ldots, n$ are the complex amplitudes of the fields incident on the corresponding ($m$-th) DOEs having the CTFs $T_{m,q}(\mat{u}_m)$, 
$C_{q,m} = ( \ii \lambda_q d_m )^{-1} \exp \{ \ii 2\pi d_m / \lambda_q \}$, and $d_m = f_m - f_{m-1}$ are the distances between the adjacent planes.

We assume that in the output plane $z = f_{n+1}$, $N_q$ spatially separated target regions $G_{q,k}, k = 1, \ldots, N_q$ are defined, which correspond to $N_q$ different classes of the problem $P_q$ (see Fig.~\ref{fig:1}).
At each input image, a certain ``energy'' distribution $E_{q,k}, k = 1, \ldots, N_q$ is generated in these regions, which corresponds to the integrals of the generated intensity distribution $I_{n+1,q,j}(\mat{u}_{n+1}) = \left| w_{n + 1,q,j}(\mat{u}_{n+1}) \right|^2$ over these regions:
\begin{equation}
\label{eq:3}
E_{q,k} = \iint I_{n+1,q,j}(\mat{u}_{n+1}) \chi_{q,k}(\mat{u}_{n+1}) \dd^2\mat{u}_{n+1} ,\,\, k = 1, \ldots, N_q,
\end{equation}
where $\chi_{q,k}(\mat{u}_{n+1})$ is the indicator function of the region $G_{q,k}$.
For solving the classification problem $P_q$, it is necessary for the cascaded DOE to generate such an intensity distribution in the output plane for the ``input signal'' of the $j$-th class $w_{0,q,j}(\mat{u}_0 )$, so that the maximum of the generated energies $E_{q,k},\,k = 1, \ldots, N_q$ is reached in the corresponding target region $G_{q,j}$~\cite{10, 12}.

Above, we described the required operation of the DNN at a single wavelength $\lambda_q$.
The problem of designing a DNN for solving several different classification problems $P_q, q = 1,\ldots, Q$ at different wavelengths $\lambda_q$ can be formulated as the problem of calculating such functions of the diffractive microrelief height $h_1(\mat{u}_1), \ldots, h_n(\mat{u}_n)$ of the cascaded DOE, so that at each wavelength $\lambda_q$ and an input signal being an image of a certain object of the problem $P_q$ solved at this wavelength, the DNN provides the maximum energy in the target region corresponding to the class of the input image.

\section{Gradient method for designing spectral DNNs}
For solving the described problem of calculating a spectral DNN, we will use a stochastic gradient descent method in the way, in which it is typically used for training artificial neural networks.
Let us first present a general description of the method.
We assume that for the calculation (training) of the DNN (cascaded DOE), a training set $S = S_1\cup\cdots\cup S_Q$ is used, which consists of training subsets $S_q$ for the considered classification problems $P_q, q = 1, \ldots, Q$.
Each training set $S_q$ contains a number of input distributions (complex amplitudes of the fields) generated from the images of the objects of the problem $P_q$ at the wavelength $\lambda_q$.
At each step of the method, a set of distributions (referred to as a batch) is randomly chosen from the training set $S$. 
For this batch, we calculate the derivatives of a certain error functional $\varepsilon ( h_1, \ldots, h_n)$, which depends on the functions of the diffractive microrelief height and evaluates the DNN performance.
Then, a step in the direction of the anti-gradient is performed, which gives the updated microrelief heights. 
Since the mathematical expectations of the derivatives calculated over a batch are proportional to the derivatives of the functional calculated for the whole training set, such an approach corresponds to the stochastic gradient descent method.
Without loss of generality, we will assume that the batch corresponds to the following set of input distributions: $w_{0,q,j}(\mat{u}_0 ),\,\,q = 1,\ldots, Q,\,\,j = 1, \ldots, N_q$.
Thus, we assume that the batch contains $N_1 + N_2 + \ldots + N_Q$ input distributions, and for each $q \in \{ 1,\ldots ,Q \}$, it includes $N_q$ images of the objects of different classes from the training set $S_q$ generated at the corresponding wavelength $\lambda_q$.
In order to describe the calculations carried out for the batch, let us write the error functional in an explicit form.
Let the classification error of an incident beam $w_{0,q,j}(\mat{u}_0)$ representing an object from the $j$-th class from the problem $P_q$ be described by a certain error functional $\varepsilon_{q,j}( h_1, \ldots, h_n )$.
Since the classification is carried out by analyzing the energies $E_{q,k}$ in the regions $G_{q,k}$ [see Eq.~\eqref{eq:3}], the functional $\varepsilon_{q,j} ( h_1, \ldots, h_n )$ in the general case has the form
\begin{equation}
\label{eq:4}
\varepsilon_{q,j}( h_1, \ldots, h_n ) = D_{q,j}(E_{q,1}, \ldots, E_{q,{N_q}}),
\end{equation}
where $D_{q,j}$ is a certain function describing the deviation of the generated energy distribution~\eqref{eq:3} from the required distribution, in which the energy is concentrated in the required $j$-th target region.
Then, the error functional for a batch containing the distributions $w_{0,q,j}( \mat{u}_0 ),\,\,q = 1,\ldots ,Q,\,\,j = 1,\ldots, N_q$ can be represented as a sum of the presented functionals:
\begin{equation}
\label{eq:5}
\varepsilon ( h_1, \ldots, h_n ) = \sum_{q = 1}^Q \sum\limits_{j = 1}^{N_q} \varepsilon_{q,j}( h_1, \ldots, h_n ) .
\end{equation}

For the functional~\eqref{eq:5}, it is easy to find the Fr{\'e}chet derivatives $\delta \varepsilon / \delta h_m$.
Indeed, since the functional~\eqref{eq:5} is equal to the sum of functionals, its derivatives have the form
\begin{equation}
\label{eq:6}
\frac{\delta \varepsilon ( h_1, \ldots, h_n)}{\delta h_m} 
= 
\sum_{q = 1}^Q \sum_{j = 1}^{N_q} \frac{\delta \varepsilon_{q,j}( h_1, \ldots, h_n )}{\delta h_m} 
,\,\,m = 1,\ldots ,n.
\end{equation}

Let us consider the calculation of the derivative $\delta \varepsilon_{q,j} / \delta h_m$ in Eq.~\eqref{eq:6} with respect to the function $h_m$.
To do this, let us first denote by
\begin{equation}
\label{eq:7}
\Delta_m \varepsilon_{q,j} ( h_1, \ldots, h_n ) = \varepsilon_{q,j}( h_1, \ldots, h_m + \Delta h_m, \ldots, h_n ) - \varepsilon_{q,j}( h_1, \ldots, h_m, \ldots, h_n )
\end{equation}
the increment of this functional caused by an increment $\Delta h_m$ of the microrelief height function ${h_m}$.
According to Eqs.~\eqref{eq:4} and~\eqref{eq:3}, this increment reads as
\begin{equation}
\label{eq:8}
\begin{aligned}
 & \Delta_m \varepsilon_{q,j} ( h_1, \ldots, h_n ) = \sum_{k = 1}^{N_q} \frac{\partial D_{q,j}}{\partial E_{q,k}}(\Delta_m E_{q,k})
\\&=
\sum_{k = 1}^{N_q} \frac{\partial D_{q,j}}{\partial E_{q,k}}\iint \left[ \Delta_m I_{n + 1,q,j}(\mat{u}_{n+1} ) \right]\cdot \chi_{q,k} (\mat{u}_{n+1}) \dd^2\mat{u}_{n+1}
\\&=
\sum_{k = 1}^{N_q} \frac{\partial D_{q,j}}{\partial E_{q,k}}\iint \Delta_m\left[ w_{n + 1,q,j}(\mat{u}_{n+1} )w_{n + 1,q,j}^*(\mat{u}_{n+1} ) \right]\cdot {\chi_{q,k}}(\mat{u}_{n+1} ) \dd^2\mat{u}_{n+1}
\\&= 
2\Re  \iint \left[ \Delta_m w_{n+1,q,j}(\mat{u}_{n+1} ) \right]\cdot 
F^*_{n + 1,q,j}( \mat{u}_{n+1} )
\dd^2\mat{u}_{n+1}  
,
	\end{aligned}
\end{equation}
where $\Delta_m E_{q,k}$, $\Delta_m I_{n + 1,q,j}(\mat{u}_{n+1})$, and $\Delta_m w_{n + 1,q,j} ( \mat{u}_{n+1} )$ are the increments of the energy, intensity distribution, and complex amplitude, respectively, caused by an increment of the height $\Delta h_m$, 
and
\begin{equation}
\label{eq:9}
F_{n + 1,q,j}( \mat{u}_{n+1} ) = w_{n + 1,q,j}(\mat{u}_{n+1} )\cdot \sum_{k = 1}^{N_q} \chi_{q,k}(\mat{u}_{n+1} )\frac{\partial D_{q,j}}{\partial E_{q,k}} .
\end{equation}
By denoting the scalar product of complex functions with angle brackets, we arrive at 
\begin{equation}
\label{eq:8b}
\Delta_m \varepsilon_{q,j} ( h_1, \ldots, h_n ) =
2\Re \left\langle  \Delta_m w_{n+1,q,j} (\mat{u}_{n+1} ), F_{n+1,q,j}(\mat{u}_{n+1}) \right\rangle.
\end{equation}

One can easily show that the operator describing the forward propagation of the light field through a set of phase DOEs [see Eq.~\eqref{eq:2}] as well as the operator of the backpropagation of the field are unitary and conserve the scalar product~\cite{16}.
Using this conservation property, we can represent the increment of the error functional~\eqref{eq:8b} as
\begin{equation}
\label{eq:10}
\Delta_m \varepsilon_{q,j} (\varphi_1, \ldots, \varphi_n) 
=
2\Re \left\langle {{{\Pr}_{{f_{n + 1}} \to f_m^+ }}({\Delta_m}{w_{n + 1,q,j}}),{{\Pr }_{{f_{n + 1}} \to f_m^ + }}({F_{n + 1,q,j}})} \right\rangle,
\end{equation}
where $\Pr_{f_{n + 1} \to f_m^+}$ is the backpropagation operator of the field from the output plane $z = f_{n+1}$ to the plane $z = f_m^+$ located immediately after the plane of the $m$-th DOE $z = f_m$.
Note that the backpropagation of the field in the free space is described by the same Fresnel–Kirchhoff integral, in which, in contrast to the forward propagation, the propagation distance is taken with a minus sign.
The ``backward propagation'' of the beam through a phase DOE is described by the multiplication of the complex amplitude of the beam by the complex conjugate of the CTF of the DOE.
Thus, at $m = n$, the field $F_{m,q,j}( \mat{u}_{n+1} ) = \Pr_{f_{n+1} \to f_m^+ }(F_{n+1,q,j})$ has the form
\begin{equation}
\label{eq:11}
F_{n,q,j}( \mat{u}_n ) = C^*_{q,n + 1}\iint F_{n + 1,q,j}( \mat{u}_{n+1} )\exp \left\{ \ii\pi\frac{( \mat{u}_n - \mat{u}_{n+1} )^2}{\lambda_q\cdot( - {d_{n + 1}})} \right\}\dd^2\mat{u}_{n+1} .
\end{equation}
Then, at $m < n$, the field $F_{m,q,j}( \mat{u}_{n+1} )$ is calculated recursively using the following formula:
\begin{equation}
\label{eq:12}
\begin{aligned}
F_{l - 1,q,j}( \mat{u}_{l-1} ) = C^*_{q,l}\iint {F_{l,q,j}( \mat{u}_l )T_{q,j}^*(\mat{u}_l )}\exp &\left\{ \ii\pi\frac{( \mat{u}_{l-1} - \mat{u}_l )^2}{\lambda_q\cdot( -d_l)} \right\}\dd^2\mat{u}_l ,
\\&\hspace{3em}l = n,\ldots ,m + 1.
\end{aligned}
\end{equation}
Let us note that since $\Pr_{f_{n+1} \to f_m^+ }(\Delta_m w_{n + 1,q,j}) = \Delta_m(w_{m,q,j} T_{m,q})$, where $w_{m,q,j}(\mat{u}_m ) T_{m,q}(\mat{u}_m )$ is the complex amplitude of the field immediately after the plane of the $m$-th DOE upon the forward propagation, the increment~\eqref{eq:10} can be transformed as
\begin{equation}
\label{eq:13}
\begin{aligned}
\Delta_m \varepsilon_{q,j}( h_1, \ldots, h_n ) 
&= 2\Re \left\langle \Delta_m( w_{m,q,j}T_{m,q} ), F_{m,q,j} \right\rangle 
\\&= 2\Re \iint w_{m,q,j}(\mat{u}_m )\Delta {T_{m,q}}(\mat{u}_m )F_{m,q,j}^*(\mat{u}_m )\dd^2\mat{u}_m .
\end{aligned}
\end{equation}
Since
\begin{equation}
\label{eq:14}
\begin{aligned}
\Delta T_{m,q}
&= \exp \{ \ii \gamma_q(h_m + \Delta h_m) \} - \exp \{ \ii \gamma_q h_m  \} 
\\&= T_{m,q}\ii \gamma_q \Delta h_m + {\rm o}(\Delta h_m),
\end{aligned}
\end{equation}
where $\gamma_q = 2\pi [ n(\lambda_q) - 1 ] / \lambda_q$, then the principal linear part of the increment~\eqref{eq:13} can be written as the following scalar product:
\begin{equation}
\label{eq:15}
\begin{aligned}
  \delta_m \varepsilon_{q,j} ( h_1, \ldots, h_n) &=  - 2{\gamma_q}\iint \Delta {h_m}(\mat{u}_m ) \Im [w_{m,q,j}(\mat{u}_m ){T_{m,q}}(\mat{u}_m )F_{m,q,j}^*(\mat{u}_m )]\dd^2\mat{u}_m  
   \\&=
		- 2{\gamma_q}\left\langle \Delta {h_m}, \Im [w_{m,q,j} T_{m,q} F_{m,q,j}^*] \right\rangle . 
		\end{aligned}
\end{equation}

According to Eq.~\eqref{eq:15}, the Fr\'echet derivative of the functional~\eqref{eq:4} has the form
\begin{equation}
\label{eq:16}
\frac{\delta \varepsilon_{q,j}( h_1, \ldots, h_n )}{\delta h_m} 
= - 2{\gamma_q}\Im [w_{m,q,j}(\mat{u}_m )T_{m,q}(\mat{u}_m )F_{m,q,j}^*(\mat{u}_m )].
\end{equation}
Thus, the calculation of the gradient of the functional for a batch can be performed using Eqs.~\eqref{eq:6} and~\eqref{eq:16}. 

Above, the functional $\varepsilon_{q,j}( h_1, \ldots, h_n )$ describing the classification error of an object of the $j$-th class in the problem $P_q$ was written in a general form~\eqref{eq:4}, where $D_{q,j}(E_{q,1}, \ldots, E_{q, N_q})$ is a certain error function depending on the energy distribution~\eqref{eq:3} generated at the functions $h_1, \ldots, h_n$.
Let us consider a particular example of the functional.
For correct classification of an input image of the $j$-th class, it is necessary for the energy $E_{q,j}$ in the corresponding region $G_{q,j}$ to have a ``large'' value $E_{\rm max}$ and for the energies in the other regions to be close to zero.
Accordingly, as an error functional for recognizing an input distribution of the $j$-th class, one can, for example, use the following quadratic functional~\cite{19}:
\begin{equation}
\label{eq:17}
\varepsilon_{q,j} ( h_1, \ldots, h_n ) = \sum_{k = 1}^{N_q} \left( E_{q,k} - E_{\rm max } \delta_{k,j} \right)^2,
\end{equation}
where ${\delta_{k,j}}$ is the Kronecker delta.
The derivatives of the functional~\eqref{eq:17} are calculated using the general formula~\eqref{eq:16}, where, according to Eq.~\eqref{eq:9}, the function ${F_{m,q,j}}\left( {\mat{u}_m } \right)$ is calculated through the backpropagation of the field
\begin{equation}
\label{eq:18}
F_{n+1,q,j} ( \mat{u}_{n+1} ) = 2 w_{n + 1,q,j}(\mat{u}_{n+1} ) \sum_{k = 1}^{N_q} \chi_{q,k}(\mat{u}_{n+1} )\cdot ( E_{q,k} - E_{\rm max }\delta_{k,j} ).
\end{equation}

Let us note that in the design of a cascaded DOE, the functions of the diffractive microrelief height $h_1(\mat{u}_1), \ldots, h_n(\mat{u}_n)$ are usually assumed to be bounded and take values from a certain interval $[0, h_{\rm max})$, where $h_{\rm max}$ is the maximum microrelief height (the $h_{\rm max}$ value is defined by the technology used for the DOE fabrication).
The presence of constraints $0 \leqslant h_m(\mat{u}_1) \leqslant h_{\rm max},\,\,i = 1,\ldots ,n$ makes the problem of designing a cascaded DOE a conditional optimization problem.
For taking into account these constraints, it is necessary to introduce the following projection operator on the set of bounded height functions into the iterative calculation process:
\begin{equation}
\label{eq:19}
{\rm P} (h) = \begin{cases}
   0,&h < 0,  \\ 
   h,&h \in [0, h_{\rm max }), \\ 
   h_{\rm max}, &h \geqslant h_{\rm max }. 
 \end{cases}
\end{equation}

In particular, the introduction of such an operator to the gradient method for designing cascaded DOEs leads to the gradient projection method, in which the height functions are updated as
\begin{equation}
\label{eq:20}
h_m^k( \mat{u}_m ) = {\rm P} \left[ h_m^{k-1}( \mat{u}_m ) - t\frac{\delta \varepsilon}{\delta h_m}( \mat{u}_m ) \right],\,\,m = 1,\ldots ,n,
\end{equation}
where the superscript $k$ denotes the iteration number and $t$ is the step of the gradient method.
Note that instead of the simplest version of the gradient method of Eq.~\eqref{eq:20}, one can utilize its various extensions, e.\,g., the widely used ADAM method~\cite{34}.

\section{Design examples of spectral DNNs}
Let us consider the calculation of a spectral DNN for solving three different classification problems $P_q, q = 1,2,3$ at the following three operating wavelengths: $\lambda_1 = 457\nm$, $\lambda_2 = 532\nm$, and $\lambda_3 = 633\nm$.
As the problems being solved, let us choose the following ones: the problem of classifying handwritten digits from the MNIST dataset at the wavelength ${\lambda_1} = 457\nm$ (problem $P_1$), the problem of classifying fashion products from the Fashion MNIST dataset at $\lambda_2 = 532\nm$ (problem $P_2$), and, finally, the problem of classifying handwritten letters from A to J (lowercase and uppercase) from the EMNIST dataset at $\lambda_3 = 633\nm$ (problem $P_3$).
Note that each of the chosen classification problems contains the objects of ten classes, i.\,e., $N_1 = N_2 = N_3 = 10$.

For the DNN design, let us use the following parameters.
We assume the input images for the classification problems $P_q, q = 1,2,3$ in the input plane to be defined on a $56 \times 56$ square grid with the step of $d = 10 \um$.
The distances between the input plane to the first DOE, between the DOEs, and from the last DOE to the output plane are the same and equal 160~mm.
The microrelief height functions in the DOE planes are defined on $512 \times 512$ square grids with the step of $10\um$.
In this case, the side length of the DOE aperture amounts to 5.12~mm.
We set the maximum height of the diffractive microrelief to be $h_{\rm max} = 6\um$.
Note that DOEs with such height can be fabricated using the standard direct laser writing technique~\cite{37, 38}.
For the sake of simplicity, as the refractive indices of the DOE material, we will use the same value $n(\lambda_1) = n(\lambda_2) = n(\lambda_3) = 1.46$, which, nevertheless, is quite close to the refractive index of fused silica at the operating wavelengths.

\subsection{Sequential solution of the classification problems}
\begin{figure}[hbt]
	\centering
		\includegraphics{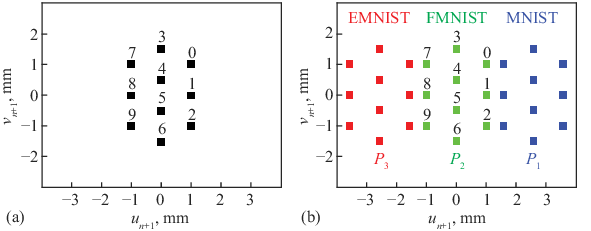}
	\caption{\label{fig:2} Target regions in the cases of sequential~(a) and parallel~(b) solution of the classification problems.}
\end{figure}

Let us first assume that the images of the objects from different classification problems $P_q, q = 1,2,3$ are generated in the input plane $z=0$ sequentially and that each image from the problem $P_q$ is illuminated by a normally incident plane wave (propagating along the $z$ axis) with the wavelength $\lambda_q$.
Since each of the considered problems $P_q$ contains the objects of 10~classes, which, as we assumed, are generated in the input plane $z=0$ in a sequential way, it is sufficient to use a single set of~10 target regions $G_k, k = 1, \ldots, 10$ in the output plane for all three problems.
In this case, the DNN will enable changing the classification problem being solved by changing the wavelength $\lambda_q$ of the incident radiation. 
The target regions, in which energy maxima have to be generated for
different classes, are shown in Fig.~\ref{fig:2}(a) and have a square shape with the side of 0.125~mm.
The classes numbered as $0,\ldots,9$ in Fig.~\ref{fig:2}(a) correspond to the digits $0,\ldots,9$ in problem $P_1$, different fashion products (T-shirt/top, trouser, pullover, dress, coat, sandal, shirt, sneaker, bag, ankle boot) in problem $P_2$, and the letters from A to J in problem $P_3$.

\begin{figure}[hbt]
	\centering
		\includegraphics{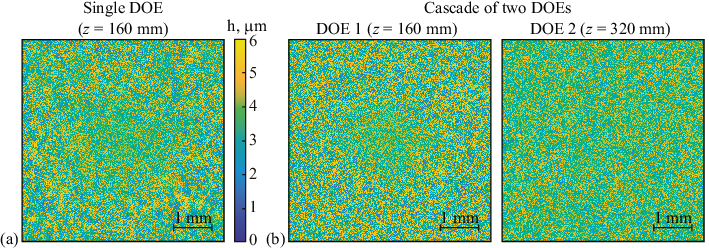}
	\caption{\label{fig:3} Microrelief height functions of the designed DNNs consisting of a single DOE~(a) and a cascade of two DOEs~(b) for sequential solution of three classification problems at three wavelengths.}
\end{figure}

First, using the developed gradient method [Eqs.~\eqref{eq:5},~\eqref{eq:6},~\eqref{eq:16}--\eqref{eq:20}], a DNN consisting of a single DOE was calculated.
Let us note that the calculation of the derivatives of the error functionals was performed numerically using the angular spectrum method~\cite{39, 40}.
For the DOE calculation, we used a training set $S$ containing 60000 images of handwritten digits from the MNIST dataset, 60000 images of fashion products from the Fashion MNIST dataset, and 48000 images of handwritten letters from the EMNIST dataset.
As the initial function of the microrelief height, a realization of a white noise with uniform distribution of the values in the $[0, h_{\rm max})$ range was used.
The training, which was carried out until reasonable convergence of the value of the error functional, took approximately 4~hours on an NVIDIA RTX 3060 12~Gb graphics card utilized for the computations.
The obtained microrelief height function of the designed DOE is shown in Fig.~\ref{fig:3}(a).

\begin{figure}[hbt]
	\centering
		\includegraphics{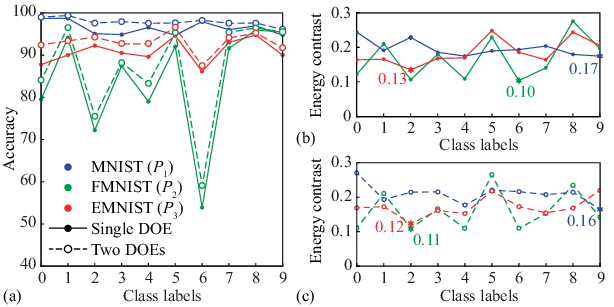}
	\caption{\label{fig:4} (a)~Classification accuracy for single-DOE (solid lines) and two-DOE (dashed lines) DNNs in the case of sequential solution of three classification problems at three wavelengths. (b),~(c)~Contrast for single-DOE~(b) and two-DOE~(c) DNNs. The stars show the minimum contrast values.}
\end{figure}

After the training, the performance of the calculated DOE was evaluated using a test set containing 10000 images for each of the problems $P_1$ and $P_2$ and 8000 images for problem $P_3$ (the images from the test set were not included in the training set).
The obtained values of the classification accuracies of the objects from different classes (such values are often referred to as recall) for the three considered classification problems are shown with circles in Fig.~\ref{fig:4}(a), which are connected with solid lines as a guide to the eye.
The overall classification accuracy (i.\,e., the ratio of the quantity of correctly recognized objects to the size of the test set) amounts to 96.41\% for problem $P_1$, 84.11\% for problem $P_2$, and 90.87\% for problem~$P_3$.
The full confusion matrices for the three considered classification problems describing the obtained results in more detail are given in the supplementary material (see Fig.~S1).
Let us note that a relatively low classification accuracy for the objects of the 6-th class of problem~$P_2$ (shirt) in Fig.~\ref{fig:4}(a) is caused by the fact that these objects are visually close to the objects of the classes~0,~2, and~4 (T-shirt/top, pullover, coat) (see Fig.~S1).
This effect is also present for the latter classes, albeit in this case, it is not so pronounced (see Figs.~\ref{fig:4}(a) and~S1).
Note that this feature is in agreement with the results of other works, in which this classification problem was considered~\cite{10, 17}.

In addition to the classification accuracy, another important parameter is the energy distribution in the target regions generated by the DOE.
Let us denote by $\overline{E}_{q,j\to k}$ the average energy calculated for the test set, which is directed to the $k$-th target region for the input objects of the $j$-th class from the problem $P_q$.
These average energy values are shown in Fig.~S1 in the supplementary material in the form of the so-called energy distribution matrices.
From the practical point of view, an important characteristic is the contrast value, which shows, how much the energy in the region of the class under consideration exceeds the energy in the regions corresponding to the other classes.
Let us introduce the contrast for the objects of the $j$-th class in problem $P_q$ as
\begin{equation}
\label{eq:21}
{\rm CR}_{q,j} = \frac{\overline{E}_{q,j\to j} - \max\limits_{k \ne j} \overline{E}_{q,j\to k}}{\overline{E}_{q,j\to j} + \max\limits_{k \ne j} \overline{E}_{q,j\to k}}.
\end{equation}

In the opinion of the authors, for robust identification of ``true maxima'' of the energy in the experimental implementation of the DNN, it is necessary for the theoretical values of ${\rm CR}_{q,j}$ to exceed at least~0.1.
The obtained contrast values for the three considered problems are shown in Fig.~\ref{fig:4}(b).
The minimum contrast values ${\rm CR}_{{\rm min},q} = \min_j {\rm CR}_{q,j}$ for problems $P_q, q = 1,2,3$ amount to 0.17, 0.10, and 0.13, respectively, and are not less than the chosen ``critical'' value of~0.1.

It is worth benchmarking the performance of the designed spectral single-DOE DNN solving three classification problems at three different wavelengths against separate DOEs, each of which solves a single classification problem $P_q$ at the corresponding operating wavelength $\lambda_q$.
These DOEs were calculated using the gradient method with the parameters given above.
For the calculated DOEs (not shown here for the sake of brevity), the values of the overall accuracy and minimum contrast obtained using the corresponding test set amount to~96.88\% and~0.19 (problem~$P_1$), 86.64\% and~0.11 (problem~$P_2$), and 93.3\% and 0.13 (problem $P_3$).
As one would expect, the spectral DOE [Fig.~\ref{fig:3}(a)], which enables solving all three classification problems, provides lower classification accuracies as compared to ``reference'' DOEs designed separately for each of the problems.
At the same time, the decrease in accuracy is relatively small and, for the considered problems $P_q, q = 1,2,3$, amounts to 0.47\%, 2.53\%, and 2.36\%, respectively.
The decrease in the minimum contrast for the three classification problems is also rather small.

It is also interesting to compare the performance of the calculated spectral DOE of Fig.~\ref{fig:3}(a) with the performance of a DOE solving the same three classification problems, but at a single operating wavelength.
This DOE was calculated using the gradient method for the wavelength $\lambda_1 = 457\nm$ at the parameters given above.
For this DOE (not presented for brevity), the overall accuracy and minimum contrast amount to 92.69\% and 0.12 (problem $P_1$), 81.96\% and 0.07 (problem $P_2$), and 84.9\% and 0.10 (problem $P_3$).
One can see that the single DOE solving three classification problems at the same wavelength exhibits inferior performance as compared to the spectral DOE.
The decrease in the overall classification accuracy occurring when a single operating wavelength is used instead of three different wavelengths amounts to 3.72\% (problem $P_1$), 2.15\% (problem $P_2$), and 5.97\% (problem $P_3$).
A better performance of the spectral DOE operating at three different wavelengths can be explained by the fact that the phase shifts introduced by the DOE at different wavelengths are different [see Eq.~\eqref{eq:1}].
In comparison with a DOE designed for a single working wavelength, this gives additional degrees of freedom in the optimization.

\begin{figure}[hbt]
	\centering
		\includegraphics{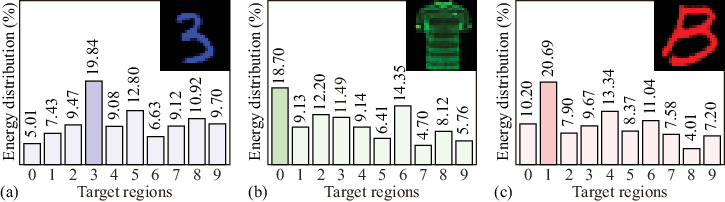}
	\caption{\label{fig:5} Examples of input images: digit ``3'' (a), object ``T-shirt/top'' (b), and letter ``B'' (c) from the classification problems $P_q, q = 1, 2, 3$ and generated energy distributions in the target regions for a DNN consisting of two DOEs.}
\end{figure}

Having discussed the properties of a single spectral DOE, let us now move to a DNN comprising two DOEs.
The microrelief height functions of the calculated DOEs are shown in Fig.~\ref{fig:3}(b).
The obtained values of the classification accuracy in the three considered problems for this DNN are shown in Fig.~\ref{fig:4}(a) with circles connected by dashed lines.
The corresponding contrast plots are shown in Fig.~\ref{fig:4}(c).
The resulting values of the overall classification accuracy and minimum contrast for the designed cascade of two DOEs equal 97.86\% and 0.16 (problem $P_1$), 86.93\% and 0.11 (problem $P_2$), and 93.07\% and 0.12 (problem $P_3$).
Full confusion matrices and energy distribution matrices for this structure are given in the supplementary material (Fig.~S2).
It is evident that the cascade of two DOEs, as compared to the single DOE, provides better performance.
In particular, the increase in the overall classification accuracy amounts to 1.45\% (problem $P_1$), 2.82\% (problem $P_2$), and 2.2\% (problem $P_3$) at virtually the same contrast.
In order to illustrate the operation of a DNN consisting of two DOEs, in Fig.~\ref{fig:5}, particular examples of input images from the classification problems $P_q, q = 1, 2, 3$ are shown, as well as the corresponding energy distributions generated by the DNN in the output plane.

In addition, we also designed a spectral DNN containing three DOEs (microrelief height functions are not shown in the paper for the sake of brevity).
For ease of comparison of the designed DNNs, Table 1 presents the values of the overall classification accuracy and minimum contrast for the single DOE and the cascades of two and three DOEs.
One can see that the values of the overall accuracy and minimum contrast for the cascade of three DOEs are 97.89\% and 0.20 (problem $P_1$), 89.75\% and 0.11 (problem $P_2$), and 93.22\% and 0.19 (problem $P_3$).
In comparison with the cascade of two DOEs, the cascade of three DOEs provides better values of the minimum contrast for problems $P_1$ and $P_3$, and a noticeably higher overall classification accuracy for problem $P_2$ (the classification accuracy increases by almost 3\%).
At the same time, the classification accuracy values for problems $P_1$, $P_3$ remain almost unchanged.

\begin{table}
\centering
\begin{tabular}{clcccccc}

\hline

  \multirow{2}{*}{\parbox{1.2cm}{\vspace{0.5em}\centering Number \\ of DOEs}}  
& \multirow{2}{*}{\parbox{1.9cm}{\vspace{1.2em}\centering Classification problem}} 
& \multirow{2}{*}{\parbox{1.8cm}{\vspace{1.2em}\centering Wavelength $\lambda$ (nm)}} 
& \multicolumn{2}{c}{Sequential regime} 
&& \multicolumn{2}{c}{Parallel regime}  \vspace{0.2em} \\ \cline{4-5}\cline{7-8} \vspace{0.2em}

& & 
& \parbox{2.0cm}{\vspace{0.5em}\centering Overall \\accuracy (\%)} 
& \parbox{1.4cm}{\vspace{0.5em}\centering Minimum \\contrast} 
&& \parbox{2.0cm}{\vspace{0.5em}\centering Overall \\accuracy (\%)} 
& \parbox{1.4cm}{\vspace{0.5em}\centering Minimum \\contrast} \\ \hline 

\multirow{3}{*}{One} 
& $P_1$: MNIST	& 457 & 96.41 & 0.17 && 96.25 & 0.18 \\
& $P_2$: FMNIST	& 532 & 84.11 & 0.10 && 83.71 & 0.11 \\
& $P_3$: EMNIST	& 633 & 90.87 & 0.13 && 90.56 & 0.14 \\ \hline

\multirow{3}{*}{Two} 
& $P_1$: MNIST   & 457 & 97.86 & 0.16 && 97.38 & 0.19 \\
& $P_2$: FMNIST  & 532 & 86.93 & 0.11 && 87.96 & 0.11 \\
& $P_3$: EMNIST  & 633 & 93.07 & 0.12 && 92.93 & 0.16 \\ \hline
 	
\multirow{3}{*}{Three} 
& $P_1$: MNIST  & 457 & 97.89 & 0.20 && 97.41 & 0.21 \\ 
& $P_2$: FMNIST & 532 & 89.75 & 0.11 && 89.10 & 0.13 \\
& $P_3$: EMNIST & 633 & 93.22 & 0.19 && 92.95 & 0.17 \\ \hline
		
		\end{tabular}
	\caption{Overall accuracy and minimum contrast provided by spectral DNNs consisting of one, two, and three DOEs solving three classification problems in sequential and parallel regimes.}
	\label{tab:1}
\end{table}

\subsection{Parallel solution of the classification problems}
In the previous subsection, we assumed that the input fields corresponding to objects from different classification problems $P_q,\,q = 1,2,3$ are generated in the input plane $z=0$ one after another, so that the DNN solves the corresponding classification problems in a sequential way.
In this case, it was sufficient to use one set of 10~target regions $G_k, k = 1, \ldots, 10$ for all three classification problems [Fig.~\ref{fig:2}(a)].
Let us now consider the case of parallel solution of the same classification problems $P_q, q = 1,2,3$.
We will assume that at each moment, three input fields with the wavelengths $\lambda_q$ are simultaneously generated in the input plane.
These fields correspond to certain objects from the considered problems of classifying handwritten digits (problem $P_1$), fashion products (problem $P_2$), and 10~handwritten letters (problem $P_3$).
Since the problems $P_q$ have to be solved simultaneously, it is necessary to define three spatially separated sets of target regions $G_{q,k}, k = 1,\ldots ,10$ corresponding to the problems being solved.
The geometry of the target regions used in the present example is shown in Fig.~\ref{fig:2}(b).

Let the images of the objects from the classification problems $P_q$ in the input plane be defined on $56 \times 56$ grids with the $d = 10\um$ step, the centers of which for different problems $P_q$ are shifted along the $u_0$ axis by different distances and are located at the points $\mat{s}_1 = (-2.56, 0)\mm$ (problem $P_1$), $\mat{s}_2 = (0,0)$ (problem $P_2$), and $\mat{s}_3 = (2.56,0)\mm$ (problem $P_3$).
These input images are schematically shown in Fig.~\ref{fig:1}.
We will assume that, in contrast to the previous case, the generated images are illuminated by obliquely incident plane waves with the wavelengths $\lambda_q$ and the propagation directions ``aimed'' from the points $\mat{s}_q$ at the center of the first DOE.
As before, in the considered DNN examples, the distances between the adjacent planes involved in the DNN design problem are the same and equal 160~mm.

\begin{figure}[hbt]
	\centering
		\includegraphics{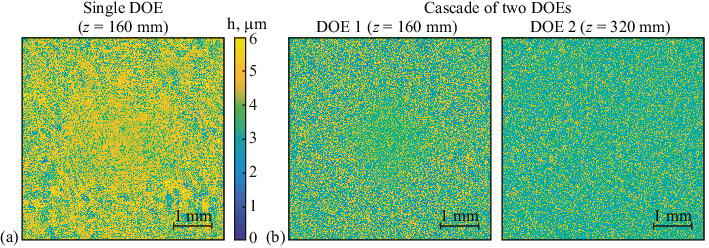}
	\caption{\label{fig:6} Microrelief height functions of the designed DNNs consisting of a single DOE~(a) and a cascade of two DOEs~(b) for parallel solution of three classification problems at three wavelengths.}
\end{figure}

\begin{figure}[hbt]
	\centering
		\includegraphics{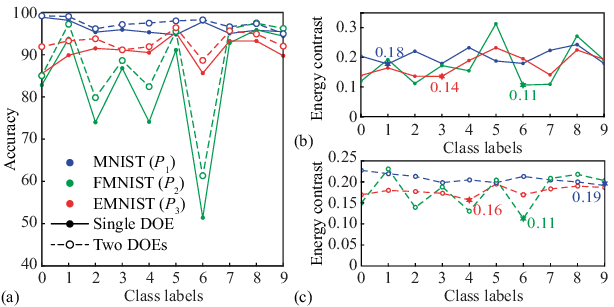}
	\caption{\label{fig:7} (a)~Classification accuracy for single-DOE (solid lines) and two-DOE (dashed lines) DNNs in the case of parallel solution of three classification problems at three wavelengths. (b),~(c)~Contrast for single-DOE~(b) and two-DOE~(c) DNNs. The stars show the minimum contrast values.}
\end{figure}

For the considered geometry of parallel solution of the classification problems using the developed gradient method of Eqs.~\eqref{eq:5},~\eqref{eq:6},~\eqref{eq:16}--\eqref{eq:20}, spectral DNN were calculated consisting of a single DOE and cascades of two and three DOEs.
As an example, Fig.~\ref{fig:6} shows the microrelief height functions of the designed single DOE and cascade of two DOEs.
In Fig.~\ref{fig:7}, the corresponding plots of the classification accuracy and contrast are shown.
The full confusion and energy distribution matrices are shown in the supplementary material in Figs.~S3 and~S4.
For ease of comparison of the performance of the designed DNNs operating in the sequential and parallel regimes, the values of the overall classification accuracy and minimum contrast for the parallel case are shown in the right part of Table~\ref{tab:1}.
By comparing Figs.~\ref{fig:4} and~\ref{fig:7} and the left and right parts of Table~\ref{tab:1}, one can see the classification accuracy values in the sequential and parallel regimes are approximately the same.
The accuracy increase ``rate'' with an increase in the number of DOEs constituting the DNN is also very similar for the sequential and parallel geometries.

Let us note that the terms ``sequential'' and ``parallel'' used for the two considered geometries are somewhat arbitrary.
In particular, the first (sequential) geometry, which corresponds to the case of normal incidence of the input beams and a single set of target regions for all classification problems being solved, can also be applied in the case of parallel processing.
In this case, similarly to work~\cite{21}, it should be assumed that in the optical setup implementing the solution of the classification problems, in addition to the DNN, there are additional optical elements that perform wavelength multiplexing of the incident beams in the input plane and wavelength demultiplexing of the resulting field distributions in the output plane.
At the same time, the second (parallel) geometry discussed in this subsection does not require the use of additional multiplexing and demultiplexing devices due to the spatial separation of input and output fields with different wavelengths.

\section{Conclusion}
We presented an approach for designing spectral DNNs (cascaded spectral DOEs) intended for solving several given classification problems at several different wavelengths, with each classification problem being solved at its ``own'' wavelength of the incident radiation.
In this approach, the problem of calculating the spectral DNN was formulated as the problem of minimizing a functional that depends on the functions of the diffraction microrelief height of the cascaded DOE and represents the error of solving the given classification problems at the design wavelengths.
Explicit and compact expressions were obtained for the derivatives of the functional and were used for formulating a gradient method for the DNN calculation.
Using the proposed method, spectral DNNs were designed for solving the following three problems: the problem of classifying handwritten digits from the MNIST database at a wavelength of 457~nm (problem $P_1$), the problem of classifying fashion products from the Fashion MNIST database at a wavelength of 532~nm (problem~$P_2$), and the problem of classifying ten handwritten letters from A to J (lowercase and uppercase) form the EMNIST database at a wavelength of 633~nm (problem $P_3$).
DNNs were designed for two geometries assuming sequential and parallel solution of different classification problems.
The presented numerical simulation results of the designed DNNs demonstrate high performance of the proposed approach.
In particular, in the parallel regime of solving the classification problems, a cascade of three DOEs provides the overall classification accuracy values of 97.41\%, 89.1\%, and 92.95\% for the problems $P_1, P_2, P_3$, respectively.

\begin{backmatter}
\bmsection{Funding}
Ministry of Science and Higher Education of the Russian Federation (State assignment to Samara University, project FSSS-2024-0016; development of a gradient method for calculating spectral DNNs and its application for solving different classification problems);
State assignment of NRC ``Kurchatov Institute'' (software development for simulating the operation of cascaded DOEs);
Russian Science Foundation (project 24-19-00080; general methodology for calculating the Fréchet derivatives of the error functionals based on the unitarity property of light propagation operators).

\end{backmatter}

\section*{Appendix. Supplementary material}

\renewcommand{\thefigure}{S\arabic{figure}}
\setcounter{figure}{0}

\begin{figure} 
	\centering
		\includegraphics{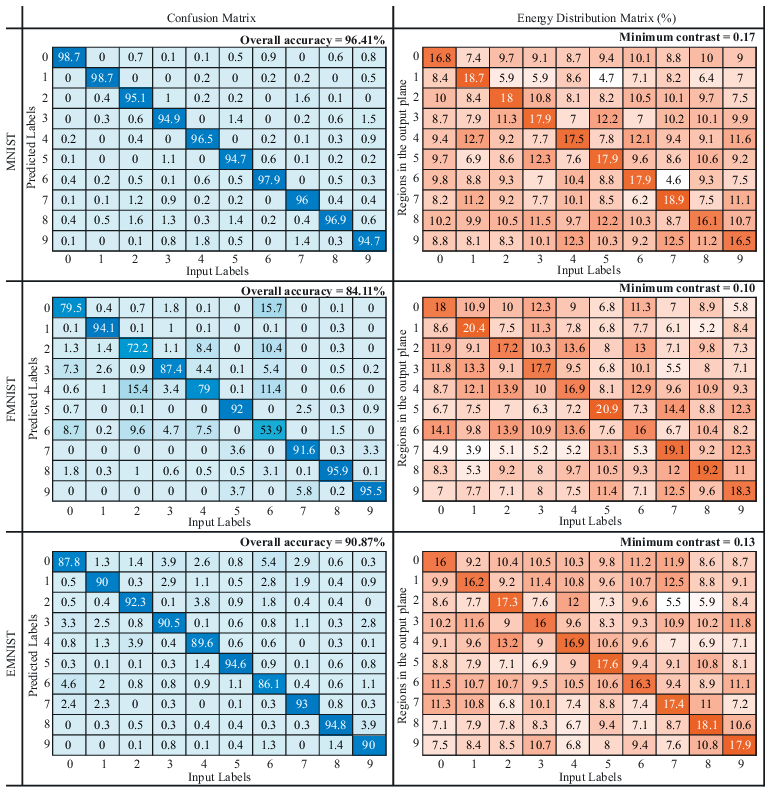}
	\caption{\label{fig:S1} Confusion and energy distribution matrices in the problems $P_1, P_2, P_3$ for a single DOE in the sequential geometry.}
\end{figure}

\begin{figure}
	\centering
		\includegraphics{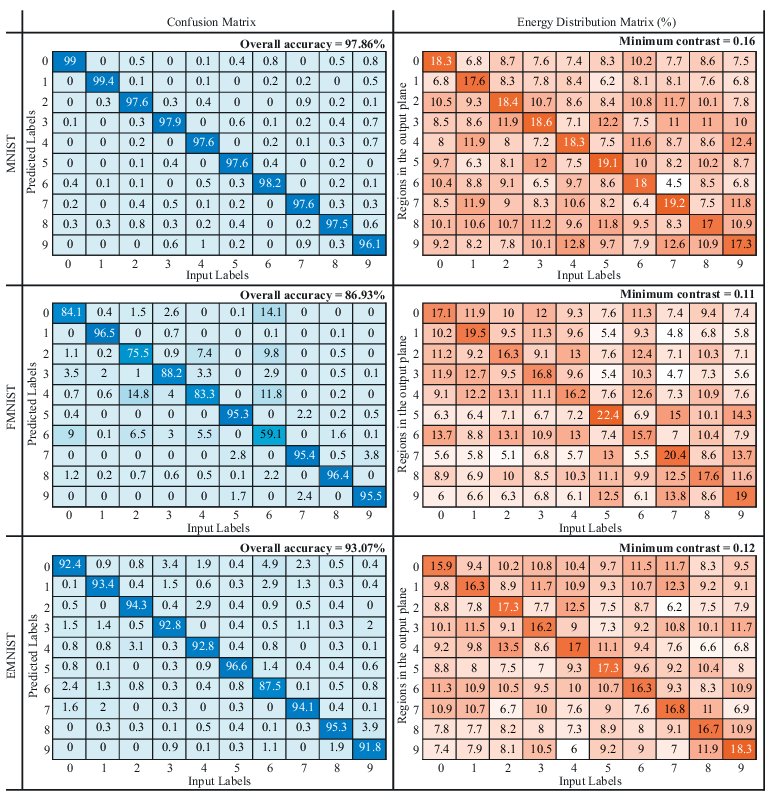}
	\caption{\label{fig:S2} Confusion and energy distribution matrices in the problems $P_1, P_2, P_3$ for a cascade of two DOEs in the sequential geometry.}
\end{figure}

\begin{figure}
	\centering
		\includegraphics{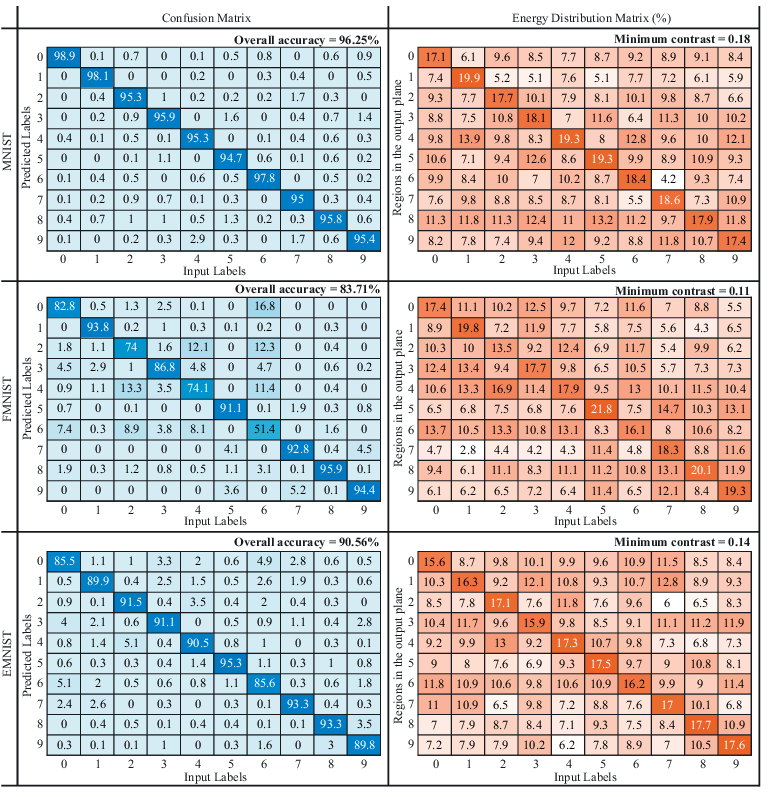}
	\caption{\label{fig:S3} Confusion and energy distribution matrices in the problems $P_1, P_2, P_3$ for a single DOE in the parallel geometry.}
\end{figure}

\begin{figure}
	\centering
		\includegraphics{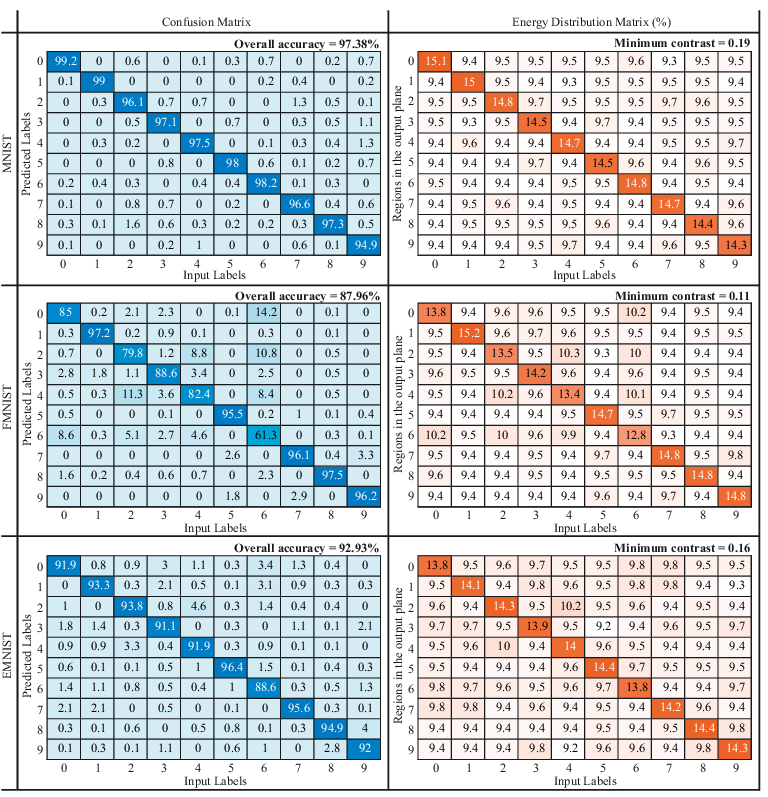}
	\caption{\label{fig:S4} Confusion and energy distribution matrices in the problems $P_1, P_2, P_3$ for a cascade of two DOEs in the parallel geometry.}
\end{figure}

Figures~\ref{fig:S1}--\ref{fig:S4} show the confusion matrices and energy distribution matrices for the designed spectral diffractive neural networks consisting of a single DOE (Figs.~\ref{fig:S1} and~\ref{fig:S3}) and a cascade of two DOEs (Figs.~\ref{fig:S2} and~\ref{fig:S4}) operating in sequential (Figs.~\ref{fig:S1} and~\ref{fig:S2}) and parallel (Figs.~\ref{fig:S3} and~\ref{fig:S4}) regimes.

\bibliography{Multiwl}

\end{document}